\documentclass{aa}  
\usepackage[dvipsnames]{xcolor}
\usepackage[breaklinks, colorlinks, citecolor=blue]{hyperref}

\usepackage{graphicx}

\usepackage{txfonts}
\begin{document} 

   \title{A VLT/VIMOS view of two {\it Planck} multiple-cluster systems: structure and galaxy properties}

   \author{R. Wicker
          \inst{1}
          ,
          N. Aghanim\inst{1}, V. Bonjean \inst{2, 3}, E. Lecoq \inst{1}, M. Douspis \inst{1}, 
          D. Burgarella \inst{3} and E. Pointecouteau \inst{4} 
          }

   \institute{Université Paris-Saclay, CNRS, Institut d’Astrophysique Spatiale, 91405, Orsay, France\\
              \email{raphael.wicker@ias.u-psud.fr}
         \and
         University of La Laguna, E-38206 Tenerife, Spain.
        Instituto de Astrofísica de Canarias, E-38205 Tenerife, Spain and University of La Laguna, E-38206 Tenerife, Spain.
         \and
             LIENSs - LIttoral ENvironnement et Sociétés (Bâtiment Marie Curie Avenue Michel Crépeau 17 042 La Rochelle cx1 - Bâtiment ILE 2, rue Olympe de Gouges 17 000 La Rochelle - France)
         \and
         Laboratoire d’Astrophysique de Marseille (UMR7326 - CNRS-INSU, Université d’Aix-Marseille)
         \and 
         IRAP, Université de Toulouse, CNRS, CNES, UPS, Toulouse, France   
             }

   \date{Received XXX; accepted XXX}

 
  \abstract  
   {We analysed spectroscopic data obtained with VLT-VIMOS for two multiple-cluster systems, PLCKG$214.6+36.9$ and PLCKG$334.8-38.0$, discovered via their thermal Sunyaev-Zel'dovich signal by \textit{Planck}. Combining the Optical spectroscopy, for the redshift determination, and photometric data from galaxy surveys (SDSS, WISE, DESI), we were able to study the structure of the two multiple-cluster systems, to determine their nature and the properties of their member galaxies. We found that the two systems are populated mainly with passive galaxies and that PLCKG$214.6+36.9$ consists of a pair of clusters at redshift $z = 0.445$ and a background isolated cluster at $z = 0.498$, whereas the system PLCKG$334.8-38.0$ is a chance association of three independent clusters at redshifts $z = 0.367$, $z =0.292$, and $z = 0.33$. 
   We also find evidence for remaining star formation activity in the highest-redshift cluster of PLCKG$214.6+36.9$, at $z = 0.498$.}

    \maketitle
%

\section{Introduction}

The largest gravitationnally bound structures in the Universe, clusters of galaxies, can often be found in pairs or in multiple systems that are identified in galaxy surveys of in X-ray observations \citep[e.g.,][]{2012A&A...539A..80L,2014A&A...567A.144C,2014MNRAS.440.1248N,2021arXiv211110253B}. These systems are of particular interest as they represent the perfect laboratories to witness the formation and evolution of the large scale structure and its impact on the baryonic matter (galaxies and gas).   

While in optical and X-ray surveys multiple cluster systems have long been detected and published in catalogues \citep[e.g.,][]{1997A&AS..123..119E,2001AJ....122.2222E}, in the (sub)millimeter surveys it is only with the achievement of the all-sky survey by \textit{Planck} \citep{2011A&A...536A...1P} between 30 and 857~GHz which allowed to detect hundreds of clusters of galaxies via the thermal Sunyaev-Zel'dovich (tSZ) effect \citep{1972CoASP...4..173S} that candidate multiple cluster systems in tSZ were observed \citep{2013A&A...550A.134P,2011A&A...536A...9P}.

The sources PLCKG$214.6+36.9$ (hereafter PLCK1) and PLCKG$334.8-38.0$ (hereafter PLCK2) are the first two triple-cluster systems discovered in the millimetre range as part of the first newly discovered tSZ sources in the \textit{Planck} early release catalogue \citep{2011A&A...536A...8P}.
The multiple nature of these systems was first shown thanks to snapshot observations with XMM-{\it Newton}, as part of the follow up program of \textit{Planck} discovered tSZ-sources  \citep{2011A&A...536A...9P}. Deeper and dedicated observation of PLCK1 et PLCK2 were later conducted in the context of a joint XMM-{\it Newton}/ESO observing program (PI: E. Pointecouteau)). A first analysis of the X-ray data for the PLCK1 system was performed by the Planck collaboration \cite{2013A&A...550A.132P}. It  allowed to identify the three components (A, B, C) forming the system and to put constraints on their redshifts from the X-ray spectroscopy of the Fe-line complex. 
The analysis indicated a redshift of $\simeq 0.45$ for the brightest component in agreement with the overall value of $z_\mathrm{PLCK1}$ but with a slight difference on the redshift of the component B was noted. 
The analysis of the deep observation of PLCK1 was revisited in \cite{lecoq2021} and Lecoq et al. (in prep).

The analysis of of the X-ray deep observation for the PLCK2 multiple-cluster system was also performed by \cite{2021A&A...653A.163K}. In addition to measuring the masses for each of the three members (A, B, C),  \cite{2021A&A...653A.163K} found that the system has an overall redshift of $z \simeq$ 0.37 but the precision of the X-ray spectroscopic analysis for two members of the triplet was found too low to rule out a projection effect along the line of sight which were identified as one of the major sources of chance associations in the construction of the \textit{Planck} cluster catalogues \citep{2014A&A...571A..29P}. 

The deeper and dedicated XMM-\textit{Newton} observation of PLCK1 et PLCK2 analysed in \cite{2013A&A...550A.132P},  \cite{2021A&A...653A.163K} and Lecoq et al. (in prep) were completed by optical spectroscopic observations from VLT/VIMOS. These optical data were destined to properly constrain the redshifts of each of the three components of the multiple systems in order to confirm or infirm their super-cluster nature and to trace the distribution of matter at the system scale. The observations were also planned so that the nature and properties of the galaxies inside the systems could be investigated. 
In this study, we present for the detailed analysis of the spectroscopic data that allow us to conclude on the nature of PLCK1 and PLCK2 sources and to deliver new results on the physical properties of their member galaxies.

We introduce the VLT/VIMOS and ancillary data in Sect. 2. We present the reduction of the VLT/VIMOS data in Sect. 3 and their analysis in Sect. 4. The results obtained on the structures of PLCK1 and PLCK2 and the properties of their member galaxies are presented in Sect. 5. Finally, we present our conclusions in Sect. 6.


\begin{table*}
\caption{Description of PLCK1 and PLCK2}             
\label{table:obs_plck12}      
\centering          
\begin{tabular}{l c c c c c c } 
\hline\hline       
Name & RA$_{SZ}$ & Dec$_{SZ}$ & RA$_X$ & Dec$_X$ & Exp. time & \# of spectra\\ 
    & [h:m:s] & [d:m:s] & [h:m:s] & [d:m:s] &  [ks] &  \\
\hline \hline
   PLCK1 & 09:08:49.44 & +14:38:31.2 & -- & -- & 28 & 5283 \\
\hline
   A & -- & -- & 09:08:49.6 & +14:38:26.8 & -- & --  \\  
   B & -- & -- & 09:09:01.8 & +14:39:45.6 & -- & --  \\
   C & -- & -- & 09:08:51.2 & +14:45:46.7 & -- & --  \\
\hline 
   PLCK2 & 20:52:42.48 & -61:12:07.2 & -- & -- & 28 & 4141 \\
\hline
   A & -- & -- & 20:52:16.8 & -61:12:29.4 & -- & --  \\  
   B & -- & -- & 20:53:08.0 & -61:10:35.3 & -- & --  \\
   C & -- & -- & 20:52:44.3 & -61:17:24.5 & -- & --  \\
\hline                  
\end{tabular}
\tablefoot{Column (1): Name of the component of the supercluster.
Columns (2) and (3) : Coordinates of the \textit{Planck} detections, from \cite{2011A&A...536A...9P}. 
Columns (4) and (5) : Coordinates of the XMM-{Newton} sources, from \cite{2013A&A...550A.132P} for PLCK1 and from \cite{2011A&A...536A...9P} for PLCK2. 
Column (6) : Exposure time for the fields of the two objects. The 28ks each are divided in four fields that have been observed 7ks each.
Column (7) : Number of individual spectra obtained after pre-processing the raw data with \textsc{Esoreflex}.}
\end{table*}

\section{Data}

\subsection{VLT/VIMOS spectroscopic data}
The PLCK1 and PLCK2 systems were observed using the VIsible Multi-Object Spectrograph (VIMOS), mounted on the Nasmyth focus B of the UT3 on the Very Large Telescope (VLT). 
The data have been acquired between 2012 and 2013, in Multi Object Spectroscopy (MOS) mode, using the Low-Resolution Red grism, with a spectral resolution R = 200 and a wavelength range covering from 5500 \AA \space to 9500 \AA. 
For each system, four overlapping fields have been observed to cover the gaps between the four quadrants of the instrument, and each field has been observed four times. Note that for PLCK1's fourth field, only three out of four observations could be used due to the lack of calibration files.

For PLCK1, all four observing fields have an overlapping region on the sky. As a result, the central region of the mosaic, i.e. the central quadrant, where components A and B lie, has been observed sixteen times in total. It is in this area that the number of galaxies is the highest. 
The masks used for the MOS being different in each field, this strategy allowed to observe a large number of nearby galaxies without risking to have their spectra overlapping.
The observing strategy for PLCK2 is slightly different, with the fields overlapping with each other so that only some regions of the global field have been observed up to twelve times. 

\subsection{Ancillary photometric data}
\label{ancillary}
In order to improve the constraints on the redshift estimate of the components of the multiple-cluster systems and to better characterise the galaxies in the field of view of PLCK1 and PLCK2, we made use of the NED\footnote{\url{https://ned.ipac.caltech.edu/}} database to complement our VLT observations.

\begin{table*}
    \centering
    \caption{Number of available spectra and galaxies after each step of the analysis of the two multiple systems.}
    \begin{tabular}{|c | c c|}
    \hline 
    {\bf System}  & {\bf \#} & {\bf Step of the analysis}\\
    \hline 
         & 5283 (VIMOS spectra) & Raw spectra \\
         & 922 (VIMOS galaxies) & After data reduction \\
         & 561 (VIMOS galaxies) & After {\sc Marz} analysis \\
         & 83 (VIMOS galaxies) & In (VIMOS)$\cap$(SDSS) \\
    PLCK1 & 27 (VIMOS galaxies) & In (VIMOS)$\cap$(SDSS)$\cap$(WISE) \\
         & 72 (VIMOS galaxies) & In (VIMOS)$\cap$(SDSS) with good {\sc Cigale} fit \\
         & 561 (VIMOS galaxies) & Used for structure analysis \\
         & 72 (VIMOS galaxies) & Used for analysis of galaxy properties \\
    \hline
         & 4141 (VIMOS spectra) & Raw spectra \\
         & 734 (VIMOS galaxies) & After data reduction \\
    PLCK2 & 299 (VIMOS galaxies) & After {\sc Marz} analysis \\
         & 2190 (DESI galaxies) & DESI galaxies in the field \\
         & 2157 (DESI galaxies) & Used for the analysis of structure and galaxy properties\\
    \hline
    \end{tabular}
    \label{tab:number_of_galaxies}
\end{table*}

The PLCK1 source is located in the north hemisphere. When available, we thus used photometric data in the five {\it u, g, r, i, z} bands of Sloan Digital Sky Survey (SDSS), namely DR16 \citep{2020ApJS..249....3A}, and the photometric data in the four AllWISE \citep{2014yCat.2328....0C} W1, W2, W3 and W4 bands. The optical photometric data will be used to estimate the stellar formation and the IR the stellar population (see Sect. \ref{SED}).

We specifically selected the galaxies from SDSS and AllWISE which were also present in the spectroscopic data from VLT/VIMOS in the field of PLCK1. To that end, we performed a cross-match based on the galaxy position between the SDSS and AllWISE catalogues, and the VLT/VIMOS sample. We associated the VLT/VIMOS sources to an SDSS or AllWISE galaxy when they fell in the same window of $5" \times 5"$, defined in the Right Ascension (R.A.) and Declination (Dec.) coordinates. 

We found SDSS photometric data for 83 of the PLCK1 galaxies. For 27 of them, AllWISE photometric data were also available. Note that in some photometric bands we had access to upper limits rather than actual flux measurements. We did not consider those.
We give in Table \ref{tab:number_of_galaxies} the summary of the number of spectra and galaxies we have access to in each step of the analysis. 

Similarly, we searched for photometric data in the direction of PLCK2 system which is located in the southern hemisphere. In this case, we used the the galaxy and galaxy-group catalogues from the Dark Energy Spectroscopic Instrument (DESI) Legacy Imaging Survey DR8 \citep{2019ApJS..242....8Z}. 
For this multiple-cluster system, we did not attempt to match the VLT/VIMOS spectroscopic sources with the DESI galaxies but rather used the latter for a structural analysis of PLCK2 (see Sect. \ref{PLCK2-desi}). We found that the DESI galaxy catalogue contained photometric redshifts, stellar masses, and star formation rates for a total of 2190 galaxies in a field of $\sim 20'$ centred around the position of PLCK2. In addition, the DESI galaxy-group catalogue contained photometric redshifts, richness and mass estimates for 506 galaxy groups in the same field of $\sim 20'$ around PLCK2. For three of these groups, we find a match with the position of the X-ray components (A, B, C) associated to PLCK2.

\section{Data Reduction}\label{reduc}

The VLT/VIMOS data were processed using the \textsc{Esoreflex} pipeline \cite[see][for detailed information and description about the pipeline]{2013A&A...559A..96F}. 
This pipeline extracts the spectra from the raw images and performs the flux and wavelength calibration using reference star spectra, and it estimates the spectrum of the sky at the moment of the observation. 

The estimation of the sky spectrum principally aims at  mitigating the atmospheric effects by removing atmospheric lines from the data. However, we have noted the presence of a remaining strong OH atmospheric line at 7600 \AA. We will see in the following (Sect. \ref{redshift}) that this residual line needs to be taken into account in the process of estimating the redshifts of the observed galaxies. It acts as a significant contamination, and source of degeneracy, when spectral features from the galaxies were not visible within the wavelength range (as was the case for PLCK2).

Eventually, the output of the {\sc Esoreflex} pipeline consists in a set of calibrated spectra and their associated coordinates. This sample includes multiply observed sources (up to 16 for the central quadrant of PLCK1), hence multiple spectra associated to the same source. The observation resulted in 5283 individual spectra for PLCK1 and 4141 for PLCK2.

The following steps of the data reduction thus aim at identifying the multiply observed spectra associated to the same source and computing a single averaged spectrum.
To do so, we have first identified the spectra of the repeated sources based on their coordinates. In practice, all sources separated by less than 1.8" in R.A. and 7.2" in Dec. were considered to be the same. 

We have then checked the overall agreement between the spectra associated with a single source, and searched for possible outliers. To do so, we computed a median spectrum, $M$, and the pixel-pixel dispersion $D$ of each spectrum with respect to the median spectrum,
\begin{equation}
    D = \sum_{i\in \mathrm{pixels}} \frac{(S_i - M_i)^2}{S_i^2},
    \label{eq:dispersion}
\end{equation}
where $S$ is the considered spectrum which is compared to the median spectrum $M$. Very similar spectra in terms of their quality and associated with the right source show low dispersion values ($\sim 1 - 10$). Outlier spectra, most likely due to an association with the wrong source, 
exhibit very high dispersion values ($> 10^{6-8}$). We find that above a threshold $D=550$ the spectrum is likely associated with the wrong source. It is hence considered for a subsequent association with another source of the sample. Finally, all the spectra associated to a multiply observed source are averaged into a single spectrum. 

At the end of the complete reduction process, we obtained 922 spectra associated with individual sources for PLCK1, and 734 for PLCK2 (see Table \ref{tab:number_of_galaxies}). 


\begin{figure*}
    \centering
        \includegraphics[width = 0.49\textwidth, trim = {2.75cm 2.25cm 0cm 5.25cm}, clip]{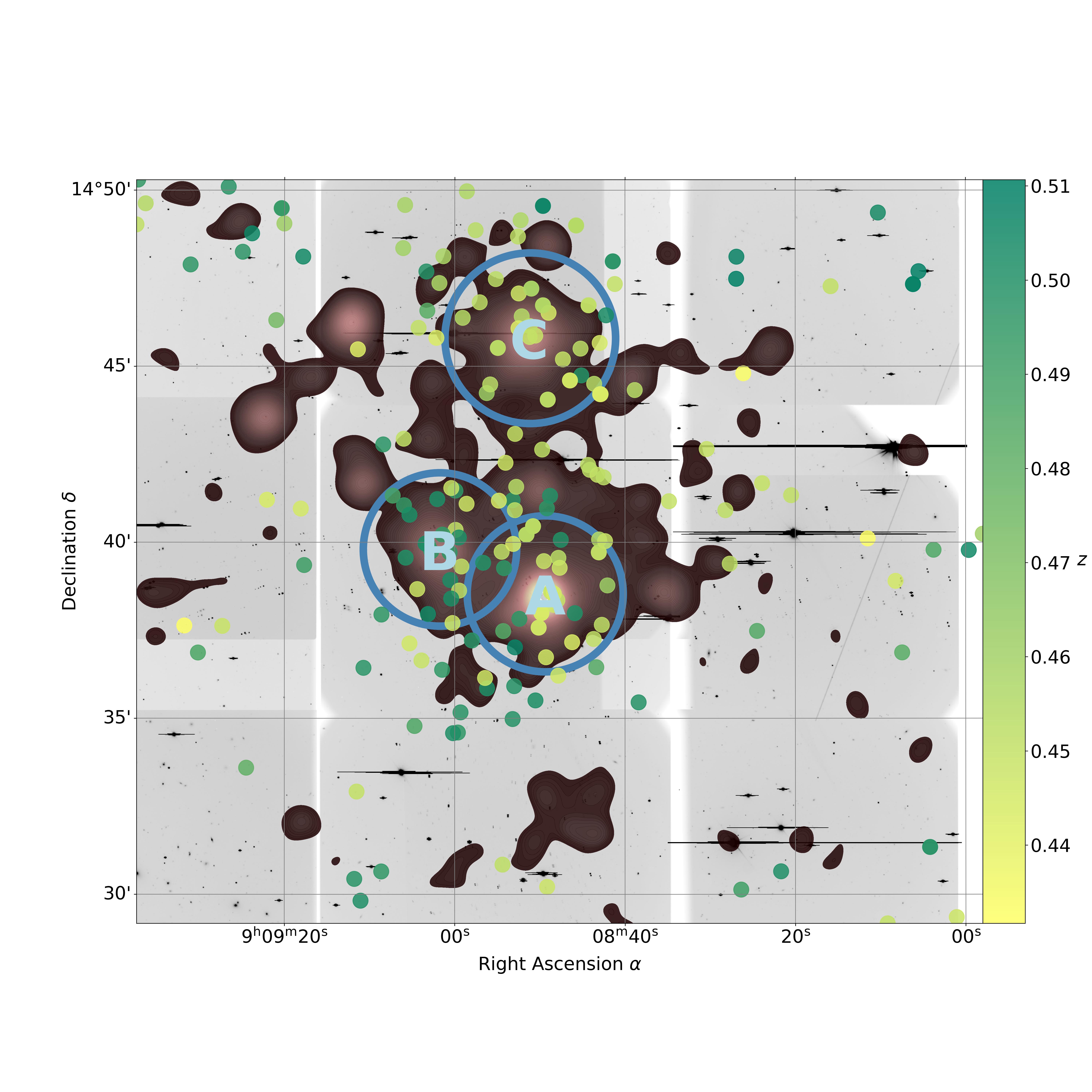}
        \includegraphics[width = 0.49\textwidth, trim={2.25cm 4.5cm 0 2.25cm}, clip]{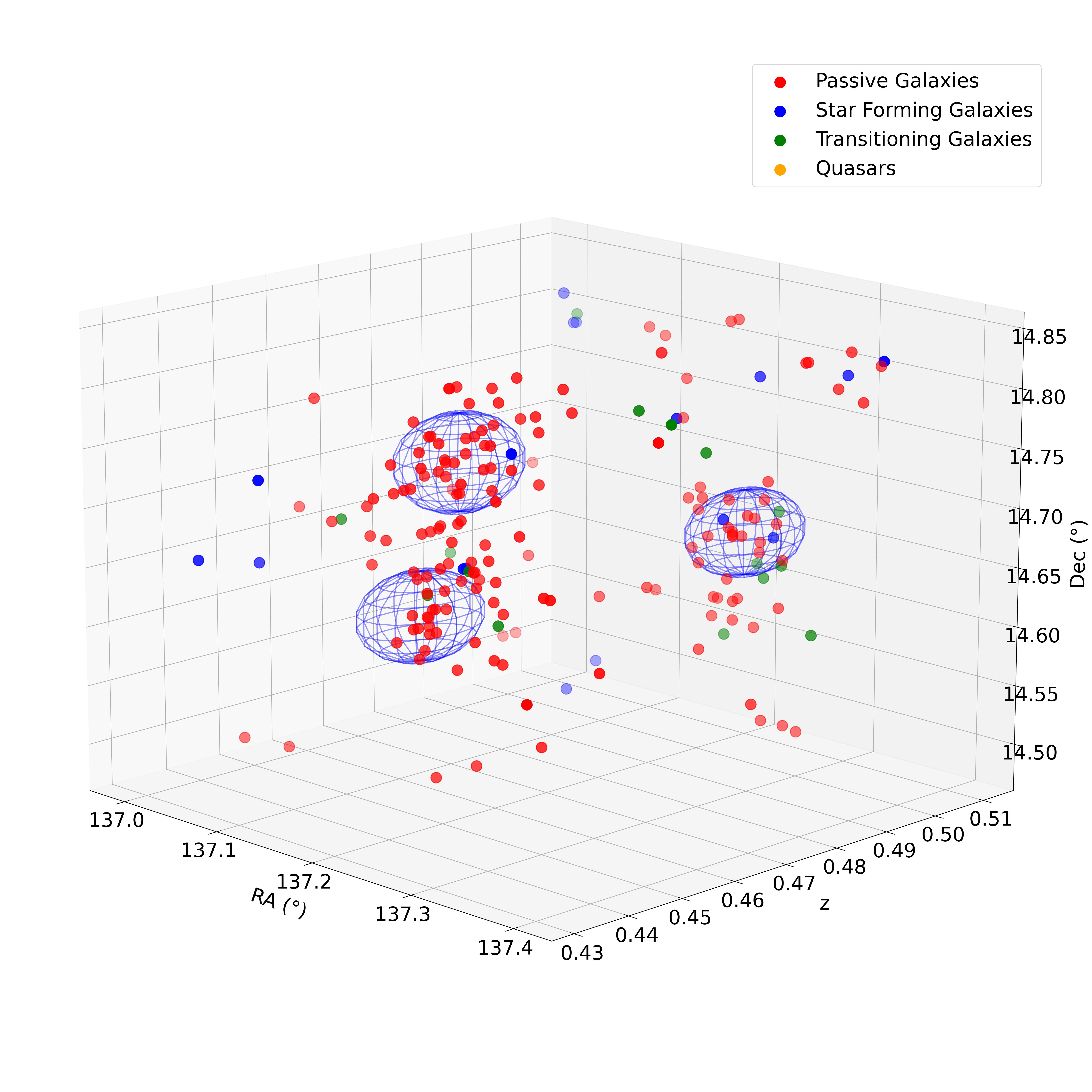}
    \caption{
    {\it Left :} VLT/VIMOS galaxies overlaid on the XMM-{\it Newton} brown-filled contours for PLCK1. The colors of the dots correspond to the redshift of the galaxies determined with {\sc Marz} (Sect. \ref{redshift}).
    {\it Right :} 3D distribution of the galaxies in the field of PLCK1 in the redshift range [0.4,0.52]. The color of the dots correspond to the galaxy types assigned with {\sc Marz} (Sect. \ref{redshift}). The blue spheres indicate the spatial extensions of the different components with radius $R_{500}$ in the R.A./Dec. plane and 3$\sigma$ in the $z$ axis.}
    \label{fig:PLCK1_redshift}
\end{figure*}

\section{Analysis}
\subsection{Determining redshifts and galaxy types from spectroscopic data}\label{redshift}

In order to analyse the reduced spectra and extract from them an estimate of the galaxy redshifts and types, we used the online tool \textsc{Marz} \citep{Hinton2016Marz}. It allows us to fit the data with either star or galaxy spectra and is based on a variety of templates which include Early Type, Star Forming or Transiting galaxy spectra, a quasar-type spectrum, and several stellar spectra (see \citep{Hinton2016Marz} for the detailed list of spectra). For each fit a "Quality Operator" (QOP) is assigned, ranging from 1 (when it is impossible to state on the quality of the fit) to 4 (when the fit is labelled as "Great") and a QOP of 6 can be assigned, when the target is a star. Eventually, the \textsc{Marz} code automatically assigns a template, a redshift and a QOP to a source, based on a correlation product between its spectrum and a template. 

We have run \textsc{Marz} on the sample of spectra for both PLCK1 (922 spectra) and PLCK2 (734 spectra) and performed a visual inspection of all the individual outputs in order to confirm the assignation of the templates, redshifts and QOP. For the multiple-cluster systems PLCK1 and PLCK2, we found 20 and 24 sources, respectively, associated with stars and discarded from the analysis. In addition and for the rest of the analysis, we discarded all the spectra for which the QOP was 1 as their nature and redshifts were not properly assessed.  

\begin{figure}
    \centering
    \includegraphics[width = 0.49\textwidth, trim={2.25cm 2.2cm 2.5cm 4.5cm}, clip]{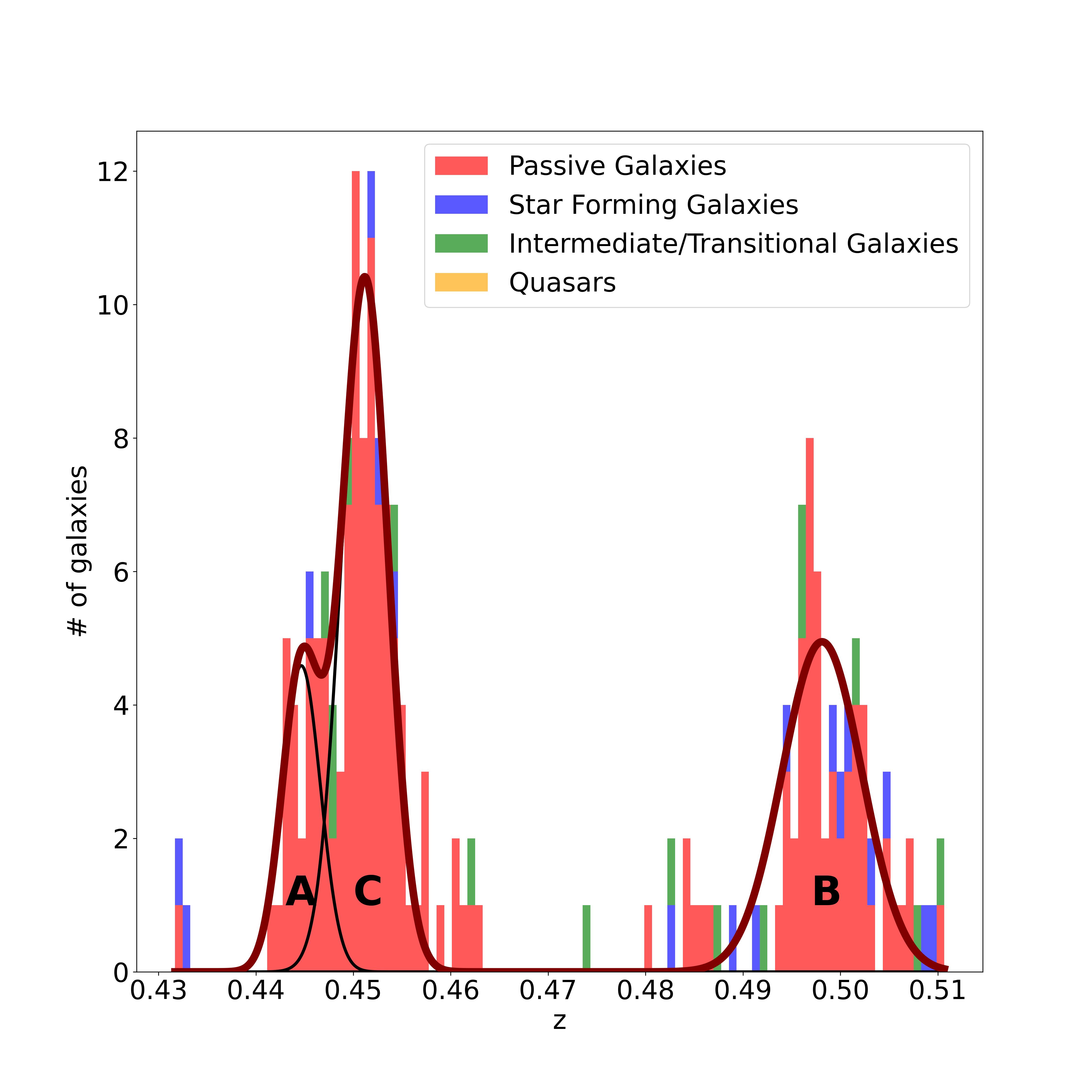}
    \caption{Cumulative histogram of the spectroscopic redshifts of 561 galaxies in the field of PLCK1. The galaxy types are based on the {\sc marz} templates used to fit the spectra (see Sect. \ref{redshift} for details). Passive galaxies are in red, star-forming ones in blue, and transitioning galaxies are in green. The solid lines show the fits to the redshift distribution in the components A, B and C of PLCK1.}
    \label{fig:plck1_histoz}
\end{figure}

It is important to note that in the case of PLCK2, with an estimated redshift $z \simeq 0.35$ from the X-ray analysis \citep{2011A&A...536A...9P}, the wavelength range of the VLT/VIMOS observation prevented us from accessing the 4000\AA \space spectral discontinuity \citep{1983ApJ...273..105B}. The 4000\AA  \space break usually serves as a great reference point for the determination of the redshift of no-emission line early type galaxies.
Combined with the contamination from atmospheric OH line at 7600\AA \space (discussed in Sect. \ref{reduc}) that perfectly matches with the Na line for an Early Type Absorption Galaxy at $z = 0.29$, the {\sc marz} code tended to automatically associate observed spectra not showing emission lines to Early Type galaxies at $z = 0.29$. Sources presenting no other recognizable spectral features than the one at 7600\AA \space had their QOP systematically set to 1 and were discarded for the rest of our analysis.

We show in Table \ref{tab:marz_results} the outputs of the this analysis step, in terms of retained spectra, for both multiple-cluster systems PLCK1 and PLCK2. Redshifts are determined for each galaxy. For the 561 galaxies retained in the PLCK1 field (Table \ref{tab:number_of_galaxies}), we find that 34.8\% of the redshifts are in the range [0.44, 0.51]; 18.2\% in of them are within [0.44, 0.455] (A and C) and 10.6\% within [0.49, 0.506] (B), the estimated cluster redshifts (see Fig. \ref{fig:plck1_histoz}). For the 299 galaxies retained in the PLCK2 field (Table \ref{tab:number_of_galaxies}), we find that 11.1\% of them are in the expected redshift range [0.345, 0.375], estimated from X-rays, and a larger fraction, 15.5\%, at $z=0.29$ (in the range [0.28, 0.3]) as expected from the contamination by the atmospheric line. 

The templates automatically assigned by the \textsc{Marz} code (see Table \ref{tab:marz_templates}) allows us to obtain galaxy types\footnote{In the following, we call transitioning galaxies the ones identified by \textsc{Marz} as intermediate, composite or transitioning.} for each of the galaxies in the observed multiple-cluster systems.  Among the 561 sources retained as galaxies in the PLCK1 field, we find that 46.1\% of them are passive, 22.4\% are transitioning and 30.4\% are star-forming galaxies (the remaining 8\% are quasars). Restricting the redshift range to that expected from the X-ray analysis of PLCK1, namely [0.44, 0.51], we find 84.3\% passive galaxies, 7.9\% transitioning and 7.9\% star-forming galaxies, indicating that the galaxy population is indeed that of clusters with quenched star formation. For the 299 galaxies of the PLCK2 field, we find 42.9\% passive galaxies, 19.9\% transitioning and 36.8\% star-forming galaxies. When restricting to the redshift range of the components A and C, namely [0.345, 0.375], we find 81.8\% passive galaxies, 3\% transitioning and 15.2\% star-forming galaxies. Here again, the population is dominated by cluster type galaxies. In the redshift range [0.28, 0.3] of the component B, we find 30.4\% passive galaxies, 34.8\% transitioning and 34.8\% star-forming galaxies. It is worth noting that the fractions are biased given that during the data reduction process we may have rejected a large number of passive galaxies $z=0.29$ due to the atmospheric contamination.

\subsection{Estimating galaxy properties from photometric data}\label{SED}

Beyond the identification of the galaxy types from the  {\sc marz} code, we investigated the galaxy properties in the two multiple-cluster systems. Namely, we estimated their stellar masses, $M_*$ and star-formation rates (SFR). 

None of the direct and main tracers of the star formation were accessible in our data either due to the spectral resolution or to the spectral range. 
Indeed, the H$\alpha$ recombination line at 6563 \AA \space and [OII] emission line at 3727 \AA, identified as good proxies for the star formation \citep{1998ARA&A..36..189K}, were not observed or blended.
Furthermore the wavelength range allows to observe the [OII] forbidden line only for galaxies with $z > 0.5$, above the estimated redshifts of PLCK1 and PLCK2 from the X-ray analysis of \cite{2011A&A...536A...9P}.
In order to constrain the the stellar properties of the galaxies, we have thus resorted to an approach based on the SED fitting and using the code {\sc Cigale} \citep{2005MNRAS.360.1413B,2009A&A...507.1793N,2019A&A...622A.103B}. 
When available, we have used the photometric fluxes from SDSS and AllWISE for the SED fitting. 
This was the case for 83 galaxies of our VLT/VIMOS sample from the PLCK1 field, for which SDSS {\it ugriz} photometric fluxes were avalaible from SDSS-DR16. 
For subset of these galaxies, AllWISE photometry in the bands W1, W2 and W3 are available (see Table \ref{tab:number_of_galaxies}). 
We eventually used this combination of data to constrain the SFR and stellar masses $M_*$ of the galaxies. 
As input models for CIGALE, we considered a delayed star formation history with optional exponential burst \citep{malek2018}, a simple stellar population (SSP) from \cite{bruzual2003} with an Initial Mass Function (IMF) from \cite{chabrier2003}, a nebular emission component, a dust attenuation law following \cite{calzetti2000}, and a dust emission following \cite{dale2014} templates. 
All models are explained in details in \cite{2019A&A...622A.103B}. The final spectra reconstructed with the combination of these components are eventually convolved with filter responses of the selected surveys to reconstruct modeled magnitudes. Hyperparameters of the models are found with best $\chi^2$ or with Bayesian methods \citep{2019A&A...622A.103B} obtained on the observed magnitudes. 
The SFR and Mstar estimated values are obtained for the best reconstructed magnitudes. 

We have visually inspected the spectra, returned by the SED fitting with {\sc Cigale}, and find a good of quality fit for 72 galaxies in total, which we use for the rest of the analysis.

\section{Results}
\subsection{PLCK1 mutliple-cluster system}
\subsubsection{Structure of the system}

We display, in Fig. \ref{fig:PLCK1_redshift} right panel, the 3D distribution of all the galaxies in the field of PLCK1. The redshifts are obtained from  {\sc marz} as explained in Sect. \ref{redshift}. The colors  passive in red, star-forming in blue, and transiting in green indicate the galaxy types based on the {\sc marz} templates. The spheres illustrate the cluster components (A, B, C) identified in \cite{2011A&A...536A...9P} with their radii (here $R_{500}$) estimated 
from the X-ray analysis of Lecoq et al. (in prep.). Figure \ref{fig:PLCK1_redshift} left panel shows the 2D projection of the galaxies, with the colour bar indicating the redshifts. We also show, in Fig. \ref{fig:plck1_histoz} the cumulative histogram distribution of galaxy types and redshifts, in the range [0.4, 0.52] where the system is expected from the X-ray redshift estimate. In this histogram, the bins contain all galaxy types. We clearly see two peaks of galaxies at $z \simeq 0.45$ and  $z \simeq 0.5$. Both correspond to concentrations of passive galaxies (in red). The first peak of galaxies matches the components A and C of PLCK1, forming a cluster pair that we call the "A-C pair" hereafter. The second peak, at $z \simeq 0.5$, matches the component B, which we call "cluster B" hereafter.

We fitted the redshift distribution for the A-C pair and the cluster B and found $z_{B} = 0.498 \pm 0.004$ ($1\sigma$) and $z_{A-C} = 0.450 \pm 0.004$ ($1\sigma$). We also fitted the redshift distribution for two individual cluster components A and C (see Fig. \ref{fig:plck1_histoz}), that gave $z_A = 0.445 \pm 0.002$  ($1\sigma$) and $z_C = 0.451 \pm 0.002$ ($1\sigma$). The spectroscopic redshifts from the VLT/VIMOS redsfhits are perfectly compatible the redshifts determined from the re-analysis of the XMM-{\it Newton} data in Lecoq et al (in prep.) (see Table \ref{tab:PLCK_structure}). Assuming the cosmological parameters from \cite{2020A&A...641A...6P}, the difference in redshift between A-C pair and B components, $\Delta z = 0.048 \pm 0.008$, corresponds to a comoving distance of $d = 163 \pm 28$ Mpc. The PLCK1 multiple-cluster system that was detected by \textit{Planck} is thus the combined signal from the projection of a cluster pair and an isolated background cluster.

\begin{table}[h!]
    \centering
    \begin{tabular}{ccccc}
    \hline
     &  & $z_{spec}$ (This work) & $z_X$ &  \\
    \hline 
    & A     &  0.445 $\pm$ 0.002 & 0.450 $\pm$ 0.01 &\\
    PLCK1 & B     &  0.498 $\pm$ 0.004 & 0.500 $\pm$ 0.02 & \\
    &C     &  0.451 $\pm$ 0.002 & 0.450 $\pm$ 0.01 &\\
    \hline
    \hline
     &  & $z_{spec}$ (This work) & $z_X$ & $z_{phot}$ (DESI) \\
    \hline 
    & A     &  0.367 $\pm$ 0.003  & 0.37 $\pm$ 0.01 & 0.39\\
    PLCK2 & B &  0.292 $\pm$ 0.002 & 0.27 $\pm$ 0.02 & 0.29\\
    & C     &  -- & $0.33^{+0.04}_{-0.05}$ & 0.34\\
    \hline
    \end{tabular}
    \caption{{\it Top} : Redshift of the PLCK1 components, in optical spectroscopy and in X-ray spectroscopy from Lecoq et al. in prep. {\it Bottom} : Redshift of the components of PLCK2, in optical spectroscopy and in X-ray spectroscopy from \cite{2021A&A...653A.163K} and from the DESI photometric redshifts.}
    \label{tab:PLCK_structure}
\end{table}

\subsubsection{Galaxy properties}\
\begin{figure}
    \centering
    \includegraphics[width = 0.5\textwidth, trim={0 0.5cm 1.5cm 3cm}, clip]{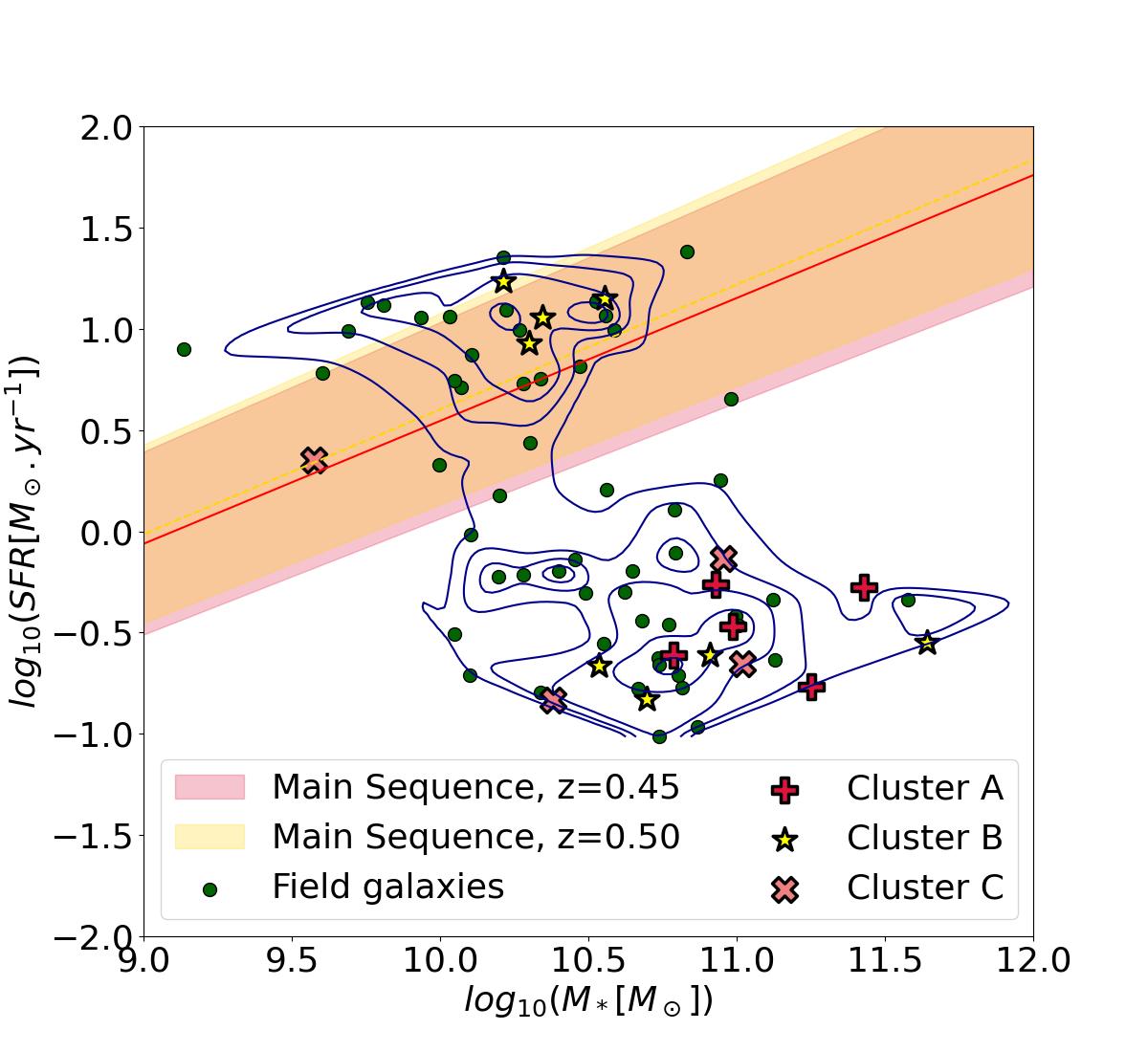}
    \caption{Distribution of the 72 PLCK1 galaxies with photometric measurements from SDSS and AllWISE on the SFR/$M_*$ plane (blue contours). Member galaxies of the A (pluses), B (stars), and C (crosses) components are identified as the galaxies within $R_{500}$ from the cluster centers and within $3\sigma$ in redshift. Field galaxies are shown as green dots. Coloured areas represent the main sequence at $z=0.45$ (A-C redshift) and $z=0.5$ (B redshift). }
    \label{fig:PLCK1_sfr_mstar}
\end{figure}
\begin{figure*}
    \centering
    \includegraphics[width=\textwidth, trim={3cm 0.5cm 0.25cm 3cm}, clip]{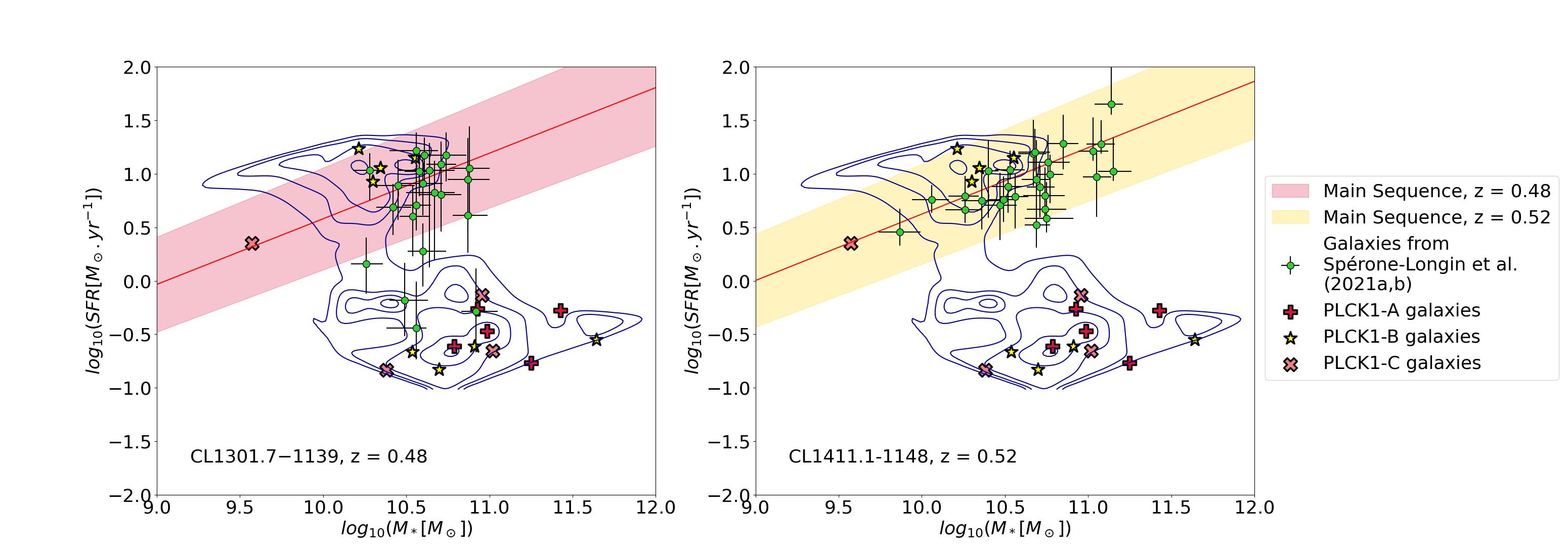}
        \caption{{\it Left} : Comparison between the stellar properties of the galaxies in PLCK1 and those in CL1301.7-1139, at $z = 0.48$, from \cite{2021A&A...654A..69S}. The red shaded area marks the galaxy Main Sequence at $z=0.48$. {\it Right} : Comparison between the stellar properties of galaxies in PLCK1, and those in CL1411.1-1148, at $z = 0.52$, from \cite{2021A&A...647A.156S}. The yellow shaded area marks the galaxy Main Sequence at $z=0.52$. {\it In both figures,} The blue contours represent the photometric measurements from SDSS and AllWISE on the SFR/$M_*$ plane. Member galaxies of the A, B, and C components are displayed as pluses, stars, and crosses respectively.}
    \label{fig:PLCK1_v_SperoneLongin}
\end{figure*}

\begin{figure*}
    \centering
    \includegraphics[width = 0.9\textwidth, trim = {0cm 3cm 2cm 5.5cm}, clip]{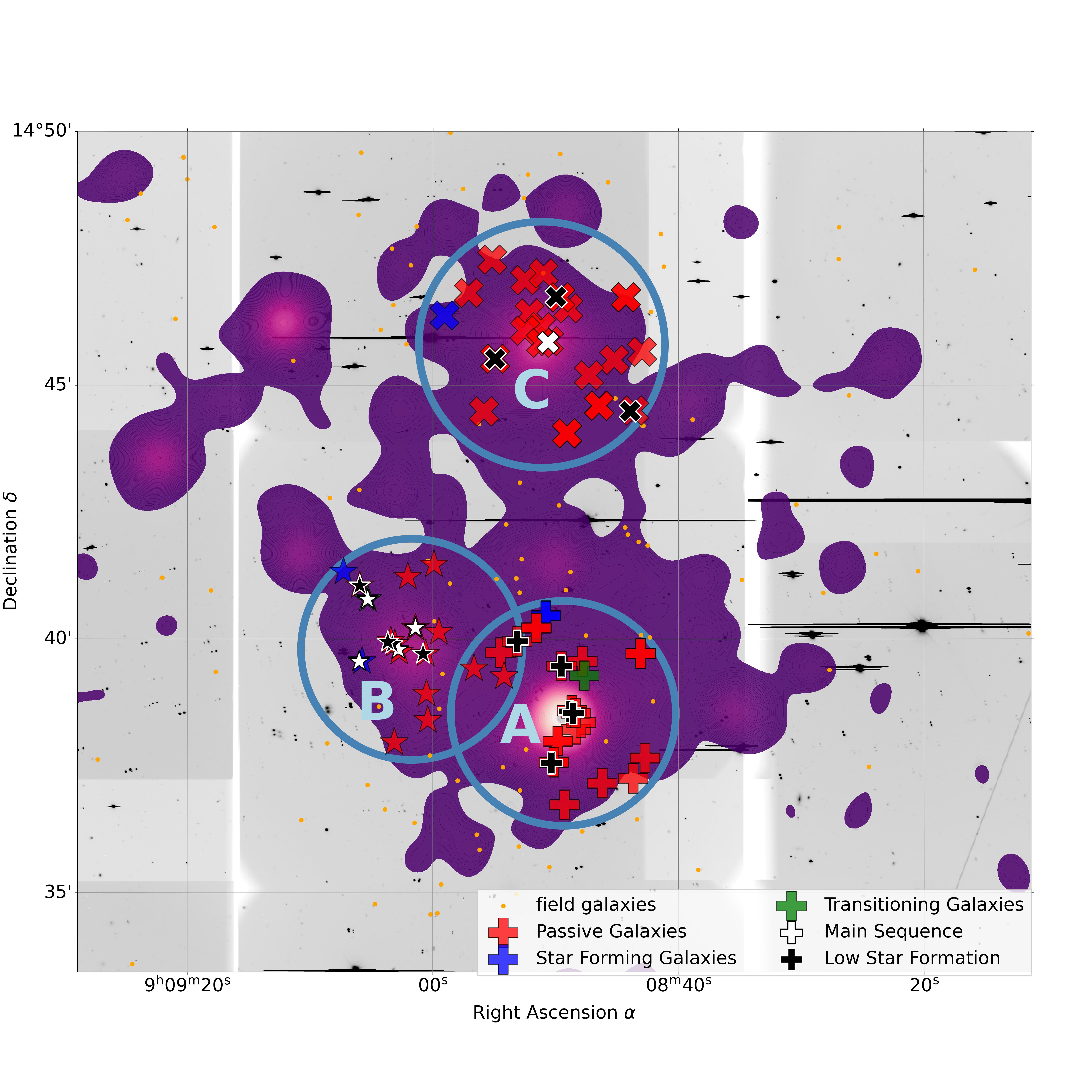}
    \caption{VIMOS galaxies overlaid on the XMM-{\it Newton} purple-filled contours for PLCK1. The "+" markers correspond to the component A, the stars to component B and "X" to component C.}
    \label{fig:plck1_optx_comp}
\end{figure*}

Based on the coordinates and redshifts of the 561 galaxies in the PLCK1 field, we determine the membership to each cluster component of the system. A cluster-member galaxy is within $3\sigma$ of the spectroscopic redshift estimate (Table \ref{tab:PLCK_structure}) and within $R_{500}$ distance from the centre of the cluster ($R_{500}$ for each cluster component is determined in Lecoq et al. (in prep)). With this definition of galaxy membership, we find that components A, B and C contain 24, 17, and 26 galaxies respectively.
Among them, five in A, eight in B, and four in C have counterparts in the SDSS and AllWISE catalogues, and had their SFR and $M_*$ determined in Sect. \ref{SED}.

We focus on the properties of these galaxy members. We display the SFR and $M_*$ obtained from the SED fitting Sect. \ref{SED} in Fig. \ref{fig:PLCK1_sfr_mstar}, where the contours represent the distribution in the PLCK1 field of the 72 galaxies with ancillary data. Figure  \ref{fig:PLCK1_sfr_mstar} clearly exhibits a bimodal distribution of the galaxies in the $\mathrm{SFR-M_*}$ plane. \\
A first population of star-forming galaxies (main-sequence galaxies hereafter) lies along a main sequence. It is compared with the computed main sequence derived from \cite{2014ApJS..214...15S} at the redshift the A-C pair, $z = 0.45$ (crimson region, in Fig \ref{fig:PLCK1_sfr_mstar}), and at the redshift of component B, $z = 0.5$ (gold region, in Fig \ref{fig:PLCK1_sfr_mstar}). A second population of galaxies consists of galaxies with low star-formation rates. In this figure, the field galaxies are represented by green dots whereas the galaxy members of the three clusters A, B, and C are shown in pluses, stars and crosses respectively. We note that most of the galaxies within $R_{500}$ from the centre of PLCK1 clusters are in the low star-forming region as expected from cluster members. This is particularly the case for the galaxies of the cluster pair A-C at $z = 0.45$ where galaxies are fully quenched. The galaxies within $R_{500}$ of the component B, at $z = 0.5$, show by contrast almost equal number of low star-forming and main-sequence galaxies indicating the star formation is still ongoing in this cluster. 
As shown in Fig \ref{fig:PLCK1_v_SperoneLongin}, this result is in agreement with the studies from \cite{2021A&A...654A..69S} and \cite{2021A&A...647A.156S} carried out at millimeter wavelengths using ALMA observations of the CO(3-2) line, within the Spatially Extended ESO Distant Cluster Survey (SEEDisCS).
Their work focuses on the galaxies between the center and the outskirts (up to $10 \times \mathrm{R}_{200}$) of the clusters CL1301.7-1139 and CL1411.1-1148, at redshifts of $z$= 0.48 and $z$ = 0.52 respectively, similar to the B cluster in PLCK1.Their study shows a dominant fraction of star-forming galaxies inside these systems.
We notice that the most distant cluster of their study, CL1411.1-1148, seems to host only main sequence star-forming galaxies, without any transitioning or passive galaxy.
The most nearby cluster of the two, CL1301.7-1139, on the contrary hosts some transitioning and passive galaxies, indicating that the process of quenching star-formation has started in this cluster.
The galaxies from the B component of PLCK1, at a redshift $z\sim 0.5$ comprised between those of the two SEEDisCS clusters, also hosts both passive and star-forming galaxies in the central region of the cluster, within $R_{500}$.
This may hint to the fact that around $z \sim 0.5$, we are witnessing the beginning of the quenching of star-formation in most individual galaxy clusters.
We find that all galaxies belonging to the main sequence are identified as "star-forming". Conversely, all galaxies, except one, identified as "Passive" are in a regime of low star formation.

It is interesting to further investigate whether there is a link between the galaxy properties and the distribution of hot gas in terms of the spatial distribution of galaxies. To do so, we display in Fig. \ref{fig:plck1_optx_comp} the member galaxies of PLCK1 overlayed on the smoothed X-ray count map (brown) and the VLT/VIMOS pre-imagery (gray quadrants). We highlight the galaxy types (passive (red symbols), star-forming (blue symbols), transitioning (green symbols)) from {\sc MARZ} and those for which we computed the stellar mass and star-formation rate with {\sc CIGALE} (white symbols for the main-sequence galaxies and black symbols for the low star-forming galaxies). 
It is worth noting that the passive galaxies are largely dominant in the A-C pair, in the cluster components, where the X-ray emissions peaks. The properties of the galaxies in the A-C pair, in terms of quenched star-formation, are similar to the case of the cluster pair A399-A401 \citep{2018A&A...609A..49B}, or to the statistical analysis of galaxy quenching inside clusters from \cite{2020A&A...635A.195G}. Finally, the galaxies laying in the main sequence (white symbols) are mainly situated in the component B which also contains low star-forming and passive galaxies.

\subsection{PLCK2 multiple-cluster system} \label{PLCK2-desi}
\subsubsection{Structure of the system}
\begin{figure*}
    \centering
    \centering
    \includegraphics[width = 0.49\textwidth, trim = {2cm 2.5cm 1.5cm 5cm}, clip]{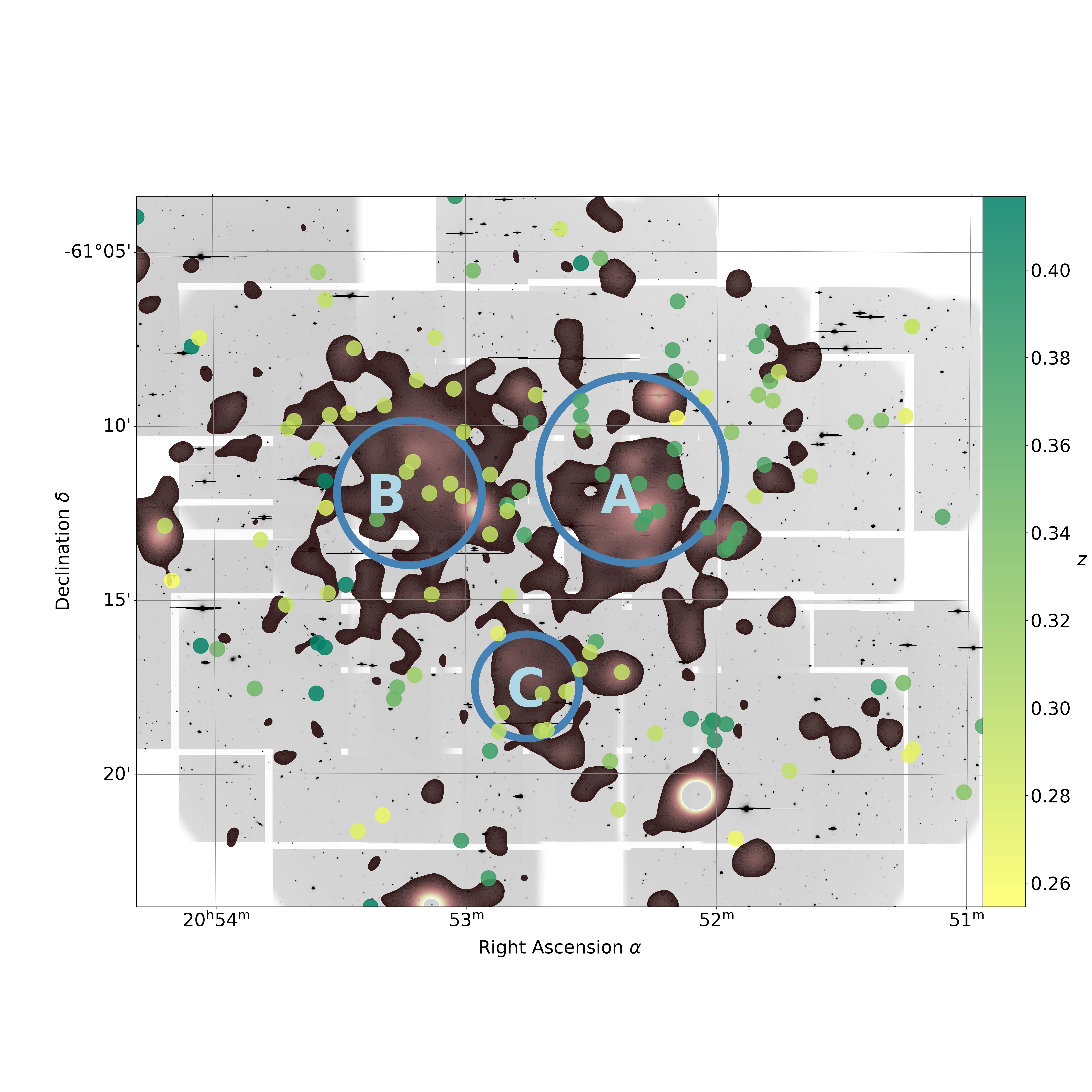}
    \includegraphics[width = 0.49\textwidth, trim = {1cm 1.5cm 1.5cm 4cm}, clip]{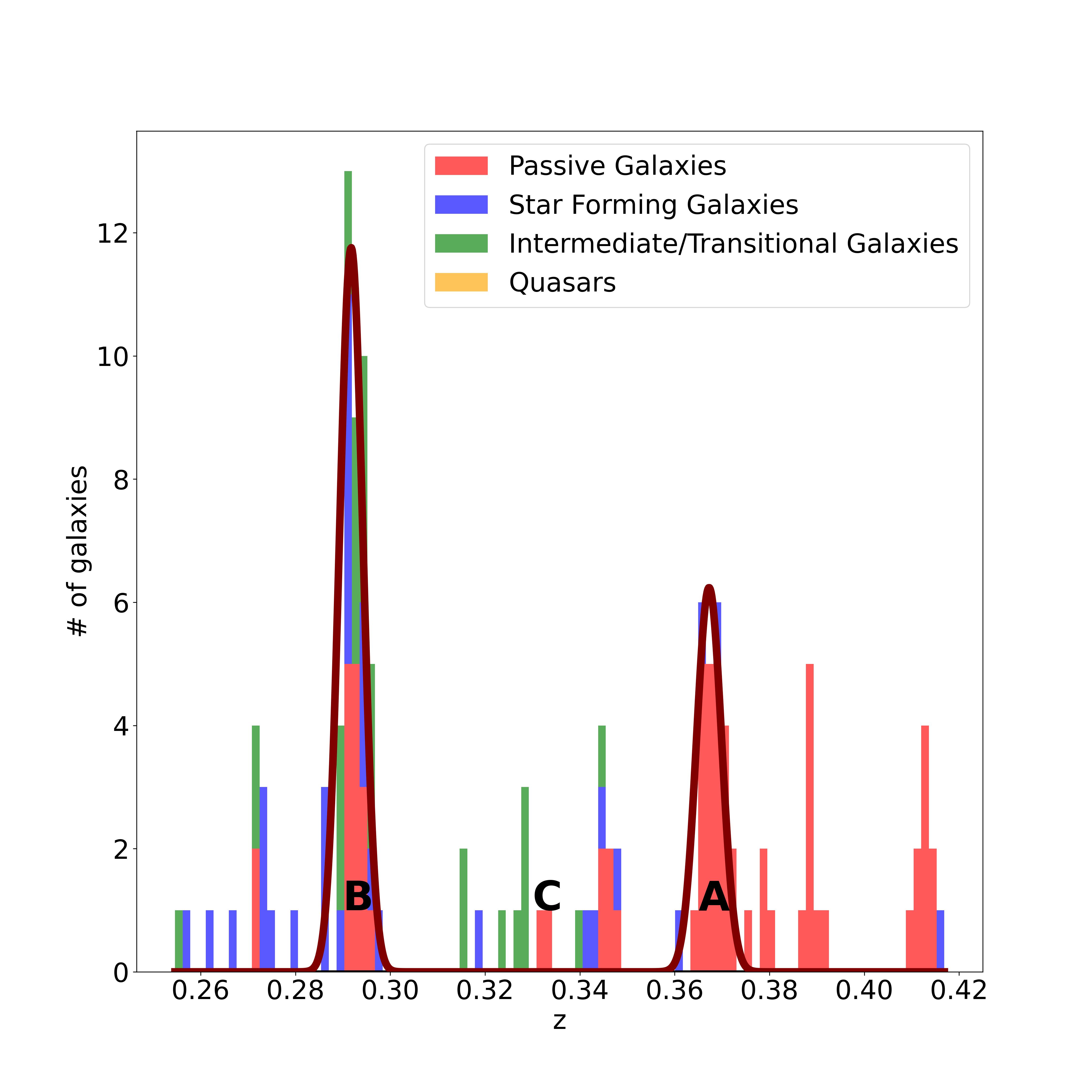}
    \caption{{\it Left :} VLT/VIMOS galaxies overlaid on the XMM-{\it Newton} brown-filled contours for PLCK2. The colors correspond to the redshift of the galaxies determined with {\sc Marz} (Sect. \ref{redshift}).    
    {\it Right : } Cumulative histogram of the spectroscopic redshifts of 299 galaxies in the field of PLCK2. The galaxy types are based on the {\sc marz} templates used to fit the spectra (see Sect. \ref{redshift} for details). Passive galaxies are in red, star-forming ones in blue, and transitioning galaxies are in green. The solid lines show the fits to the redshift distribution in the components A and B. No fit was obtained for component C.    }
    \label{fig:PLCK2_redshift}
\end{figure*}

\begin{figure}
    \centering
    \includegraphics[width = 0.5\textwidth, trim = {0cm 3cm 0cm 6cm}, clip]{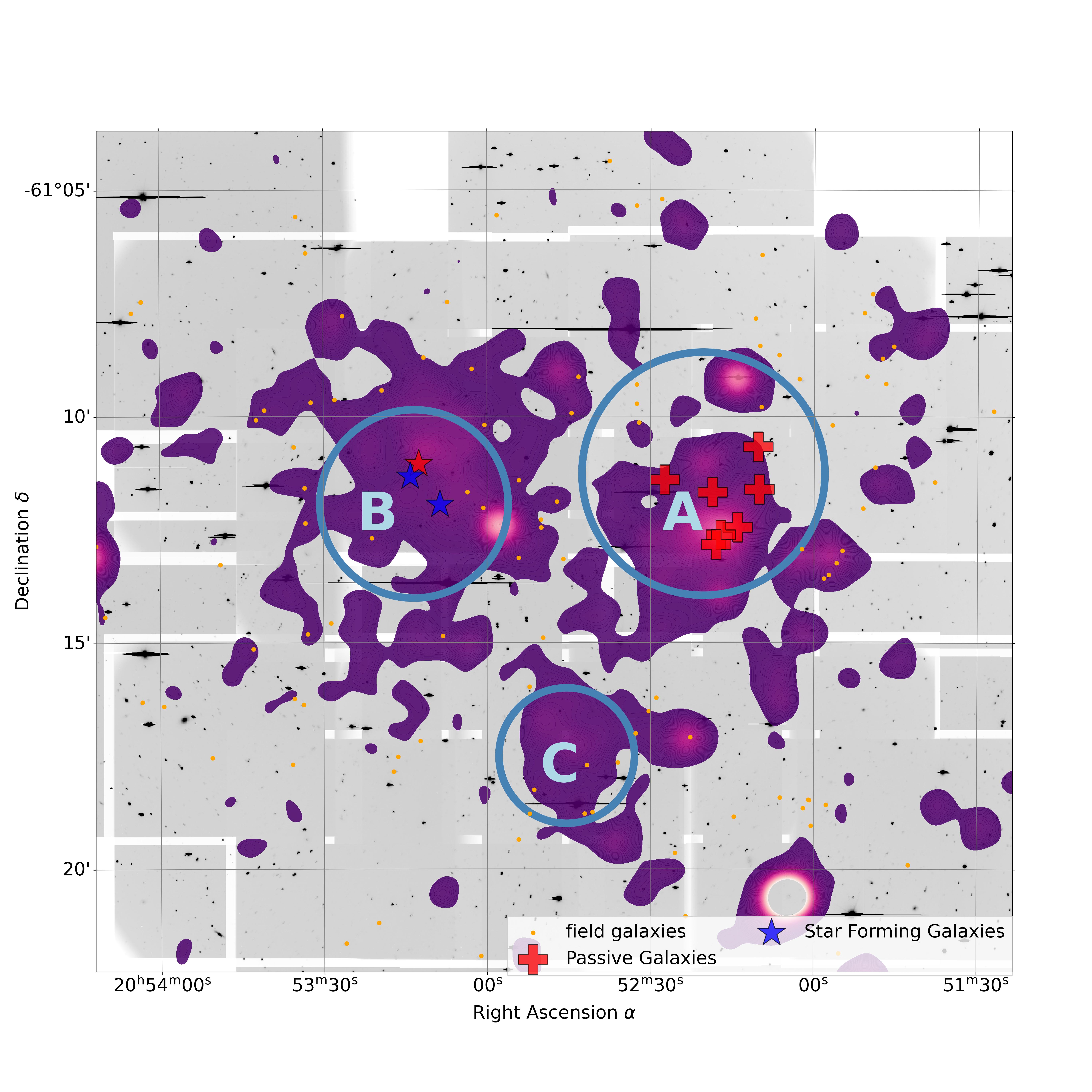}
    \caption{VIMOS galaxies overlaid on the XMM-{\it Newton} purple-filled contours for PLCK2. The "+" markers correspond to the component A, the stars to component B and "X" to component C.}
    \label{fig:plck2_optx_comp}
\end{figure}

\begin{figure*}
    \centering
    \includegraphics[width = \textwidth]{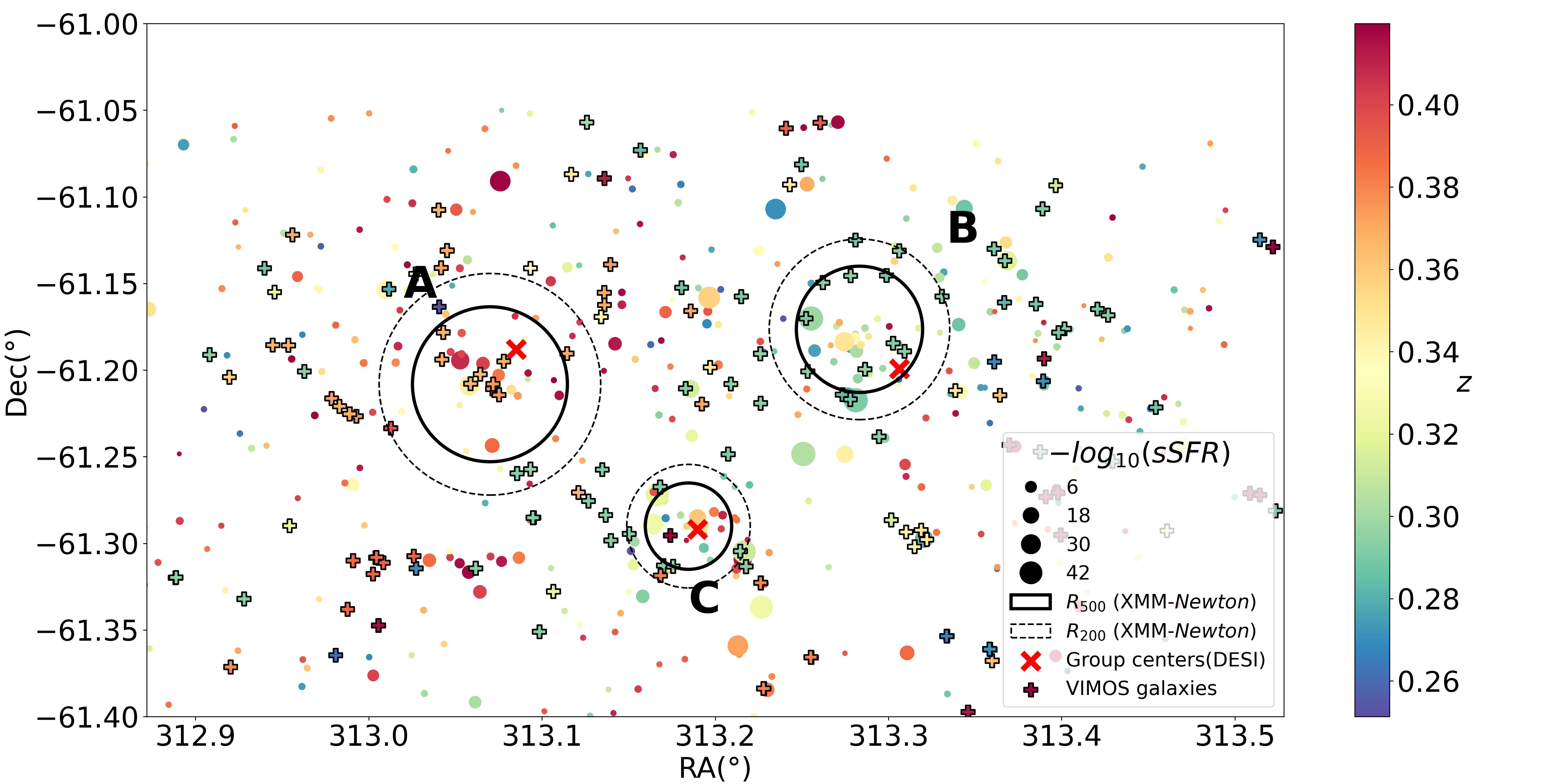}
    \caption{DESI and VLT/VIMOS galaxies inside the PLCK2 field, between $z = 0.25$ and $z = 0.42$. The size of the dots correspond to the sSFR and their color to the redshift. }
    \label{fig:PLCK2_in_DESI_withspec}
\end{figure*}

\begin{figure}
    \centering
    \includegraphics[width = 0.5\textwidth]{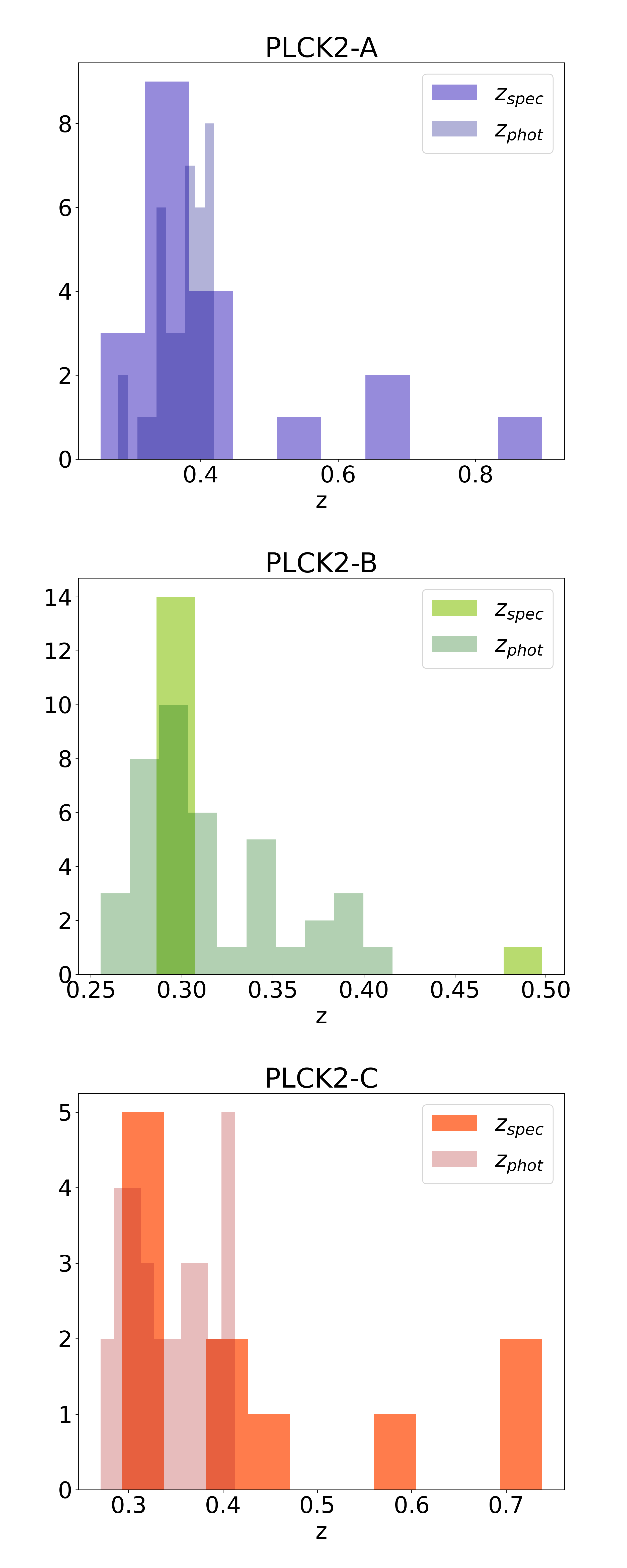}
    \caption{Comparison of the spectroscopic redshifts from VLT/VIMOS and photometric redshsifts from DESI within the R$_{500}$ of components A, B and C of PLCK2.}
    \label{fig:PLCK2_in_DESI_zspec_zphot}
\end{figure}

As discussed in Sect. \ref{redshift}, the VLT/VIMOS spectroscopic data in the field of PLCK2 suffered two main issues which limited our capacity to determine high quality and large number of redshifts for the components A, B and C. The wavelength range was not quite adapted to the expected redshift of the PLCK2 system and the contamination from the OH atmospheric line introduced a degeneracy with sources at $z=0.29$ which translated into rejecting a large number of potentially actual galaxies.

We display, in Fig. \ref{fig:PLCK2_redshift} right panel, the cumulative distribution of redshifts obtained from the analysis the  data in the PLCK2 field with {\sc marz} in the range [0.25, 0.42]. As in Fig. \ref{fig:plck1_histoz}, the colours indicate the galaxy types (passive in red, star-forming in blue, transitioning in green and quasars in orange) assigned to the sources in the PLCK2 field. We clearly see a large peak of star-forming and transitioning galaxies in the redshift distribution at $z=0.29$. Their high relative abundance with respect to passive galaxies clearly results from the contamination by the atmospheric line. We also note that all other peaks in the distribution are constituted of a handful galaxies each, most of them passive galaxies. The largest of them shows a concentration of galaxies at $z\simeq 0.37$. 

The galaxies in PLCK2 field and in the redshift range [0.25, 0.42] are displayed in Fig. \ref{fig:PLCK2_redshift} (left panel) with the VLT/VIMOS pre-imaging image (gray) and smoothed X-ray contour map (brown) displayed in the background. The blue circles represent the cluster sizes, with their radii $R_{500}$ estimated via the analysis of the XMM-\textit{Newton} data by \cite{2021A&A...653A.163K} and the colour bar indicating the redshift.  \\
We fitted the redshift distribution for the components A and B only, the peak at the expected redshift of component C being not significant. We estimated the redshift for cluster A as $z_{A} = 0.367 \pm 0.003$ ($1\sigma$) in perfect agreement with the redshift estimated from the X-ray analysis of \cite{2021A&A...653A.163K} and in good agreement with the photometric redshift from DESI. For cluster B, we found $z_{B} = 0.292 \pm 0.002$ ($1\sigma$) in good agreement with both the X-ray and photometric redshifts (see Table \ref{tab:PLCK_structure}). 

Given the lack of spectroscopic redshifts to determine the redshift of component C and hence to confirm or infirm the nature of PLCK2, we resorted to ancillary data presented in Sect. \ref{ancillary}. First, we cross-matched the sources from the VLT/VIMOS analysis with the DESI galaxy catalogue. Second, we matched the position of the DESI catalogue of groups and clusters with the positions of the PLCK2 components. In both cases, we have performed the positional matching within a region of $R_{500}$ radius, the values for A, B and C being estimated in \cite{2021A&A...653A.163K}. We illustrate the output of this cross-match in Fig. \ref{fig:PLCK2_in_DESI_withspec}. The black solid circles around A, B and C components are the $R_{500}$ circular areas whereas the black dashed lines represent $R_{200}$. The red crosses are the positions of the three clusters from the DESI catalogue that were found within $R_{500}$ from PLCK2 components. The colored pluses represent the galaxies from the VLT/VIMOS analysis with $z$ in [0.25, 0.42] and the colored dots represent the DESI galaxies. 
The color of the symbols represents the redshift of the galaxies, photometric and spectroscopic.
In the case of the DESI galaxies, the size of the symbol represent the specific star formation rate $\mathrm{sSFR = SFR/M_*}$.
The cross-match of the DESI galaxies at the region of the components shows in Fig. \ref{fig:PLCK2_in_DESI_withspec} that component A contains a significant number of DESI galaxies
with compatible photometric and spectroscopic redshifts for a proper identification of a galaxy overdensity at $z\simeq 0.36-0.4$.
For component B, we found a concentration of DESI galaxies within the $R_{200}$ region at $z\simeq 0.28-0.30$, compatible with the spectroscopic redshift estimated from the VLT/VIMOS data. We show, in Figure \ref{fig:PLCK2_in_DESI_zspec_zphot}, the comparison between the photometric and spectroscopic redshift distributions within $R_{500}$ of the three components. For components A and B, we see a good match between the spectroscopic redshifts from VLT/VIMOS and the photometric redshifts from DESI.
For component C however, we find no clear evidence for a unique overdensity of galaxies at fixed redshift. 
This is highlighted in Figure \ref{fig:PLCK2_in_DESI_zspec_zphot}, where we see no particular overdensity of galaxies in the spectroscopic data.
The DESI galaxies however seem to point at a slight overdensity of galaxies between redshifts $z \sim 0.3$ and $z \sim 0.4$.
In addition, the peak of the X-ray emission matches the position of a low-mass galaxy group from the group catalog of DESI, at $z=0.33$.

The positional matching of the DESI clusters and the PLCK2 components indicates that only three clusters identified in the DESI survey are found within $R_{500}$ from A, B, and C positions with masses and photometric redshifts compatible with the ones estimated from the X-ray analysis of \cite{2021A&A...653A.163K} (see Table \ref{tab:PLCK_structure}).
The ensemble of results, that we obtained from our analysis of the VLT/VIMOS spectroscopic data combined with the ancillary data and compared with the X-ray analysis, show that the multiple-cluster system PLCK2, discovered by {\it Planck} via the SZ signal, ends up being the chance association of three independent clusters A, B and C at redshifts $z_A\simeq 0.37$, $z_B\simeq 0.27$, $z_A\simeq 0.33$ respectively.

\subsubsection{Galaxy properties}
Very few galaxies were identified as members of PLCK2 component based on the cluster sizes, estimated via the analysis of the XMM-\textit{Newton}
data, and on their redshifts. Namely, we found only seven and three galaxies with spectroscopic redshifts in components A and B respectively. No VLT/VIMOS galaxy was found in the component C within $R_{500}$ despite the presence of a concentration of galaxies in the DESI data. In Figure \ref{fig:plck2_optx_comp}, we see that all the VLT/VIMOS galaxies in cluster A are passive (red symbols) whereas a a majority of B galaxies are star-forming (blue symbols). This is due to the bias introduced during the analysis of the spectra with {\sc Marz} (see Sect. \ref{redshift}).

The limited number of VLT/VIMOS galaxies in A and B fields within $R_{500}$ from the cluster centers makes it hard to conclude firmly on the nature of galaxies in the two components. However, we can use the DESI data and particularly the provided specific star formation rates (sSFR) for each galaxy. We see in Fig. \ref{fig:PLCK2_in_DESI_withspec} overdensities of galaxies with low and very low sSFR (i.e. passive galaxies) within the $R_{200}$ of the three clusters, at redshifts matching those of the clusters. We can thus safely say that the DESI galaxies belonging to the components A and B are not main sequence galaxies but rather low star-forming or transitioning galaxies in agreement with the expected properties of cluster galaxies.

\section{Conclusions}
In this study of the two multiple-cluster systems discovered by \textit{Planck}, PLCKG214.6+36.9 and PLCK334.8-38.0, we have analysed the spectroscopic data from dedicated observations with VLT/VIMOS in combination with ancillary optical and near-IR ancillary data from SDSS, AllWISE and DESI. These observations aimed at determining the nature of the multiple-cluster systems and the properties of the their member galaxies.  

Our analysis of the spectroscopic observation of PLCKG214.6+36.9 multiple-cluster shows that the multiple-cluster system, detected in the \textit{Planck} survey via the tSZ signal, is the combined signal from the projection of a cluster pair at $z=0.445\pm 0.002$ and an isolated background cluster at $z=0.498 \pm 0.004$. This result, hinted by the analysis of \cite{2013A&A...550A.132P}, is confirmed by an independent re-analysis of the X-ray data from XMM-\textit{Newton} by Lecoq et al. (in prep).
We also show that the multiple-cluster system PLCK334.8-38.0, detected in the \textit{Planck} data via the tSZ signal, is the result of a projection effect of three independent clusters : A at $z = 0.367\pm 0.003$, B at $z=0.292\pm 0.002$ and C at $z=0.33$, aligned along the line of sight, as hinted by \cite{2021A&A...653A.163K} in their analysis the XMM-\textit{Newton} observation. 

For both PLCKG214.6+36.9 and PLCK334.8-38.0, we confirm from the ancillary data and from the VLT/VIMOS data that the member galaxies are dominated by passive low star-forming galaxies, classified either using SED fitting or with template classification. This is in agreement with the properties of cluster member galaxies. We furthermore find that the highest redshift cluster in PLCKG214.6+36.9, at $z=0.498 \pm 0.004$, is still the site of star-formation activity in agreement with studies carried out in the mm wavelength on clusters at similar redshifts \citep{2021A&A...654A..69S,2021A&A...647A.156S}. This cluster also contains transitioning galaxies as if we were witnessing the ongoing quenching of star formation.

\begin{acknowledgements}
    This research was supported by funding for the ByoPiC project from the European Research Council (ERC) under the European Union’s Horizon 2020 research and innovation program grant number ERC-2015-AdG 695561. The authors thank all members of the ByoPiC team\footnote{\url{https://byopic.eu/team}} for useful discussions. The authors acknowledge the use of the
   NASA/IPAC Extragalactic Database (NED\footnote{\url{https://ned.ipac.caltech.edu}}), funded by the National
Aeronautics and Space Administration and operated by the California Institute of
Technology. 
\end{acknowledgements}

%
%
\bibliographystyle{aa} 
\bibliography{biblio} 
\appendix
\section{Classification in {\sc marz}}
We show, in Table \ref{tab:marz_templates}, the list of templates used in the {\sc marz} tool for the classification of the sources. Table \ref{tab:marz_results} gathers the results of the template adjustment. The total number of spectra for PLCK1 and PLCK2 are given with their associated Quality Operator (QOP), and the indication of whether or not the spectra were considered for the analysis.

\begin{table*}[]
    \centering
    \caption{Summary of the templates available in {\sc marz}, from \cite{Hinton2016Marz}.}
    \begin{tabular}{cccc}
    \hline
    Template ID & Template Name & Wavelength Range & Label and color in plots\\
    \hline \hline
    1 & A Star & 3511.0 \AA $\rightarrow$ 7429.3 \AA  & -- \\
    2 & K Star & 3511.0 \AA $\rightarrow$ 7429.3 \AA & -- \\
    3 & M3 Star & 3811.7 \AA $\rightarrow$ 9192.2 \AA & -- \\
    4 & M5 Star & 3511.0 \AA $\rightarrow$ 7429.3 \AA & -- \\
    5 & G Star & 3816.8 \AA $\rightarrow$ 9200.3 \AA & -- \\
    6 & Early Type Absorption Galaxy & 2399.4 \AA $\rightarrow$ 9120.1 \AA & Passive Galaxy (Red)\\
    7 & Intermediate Type Galaxy & 2399.4 \AA $\rightarrow$ 9120.1 \AA & Intermediate/Transitioning Galaxy (Green)\\
    8 & Late Type Emission Galaxy & 2399.4 \AA $\rightarrow$ 9120.1 \AA & Star-Forming Galaxy (Blue)\\
    9 & Composite Galaxy & 3305.2 \AA $\rightarrow$ 8344.5 \AA & Intermediate/Transitioning Galaxy (Green)\\
    10 & High Redshift Star Forming Galaxy & 1800.6 \AA $\rightarrow$ 5061.4 \AA & Star-Forming Galaxy (Blue)\\
    11 & Transitional Galaxy & 3640.5 \AA $\rightarrow$ 7071.1 \AA & Intermediate/Transitioning Galaxy (Green) \\
    12 & Quasar & 900.8 \AA $\rightarrow$ 5749.6 \AA & Quasar \\
    \hline
    \end{tabular}   
    \tablefoot{The first two columns are the ID and the name of the template in {\sc marz}. The third column indicates the wavelength range of the full template, which then can be shifted when increasing the redshift. The last column represents the label and the color-code for the galaxies displayed in cumulative histograms and in the galaxy spatial distributions.}
    \label{tab:marz_templates}
\end{table*}

\begin{table*}[h!]
    \centering
    \caption{Number of spectra per QOP in each multiple-cluster system. For each line we state if the sources with a certain QOP is conserved in the rest of the analysis. }
    \begin{tabular}{cccc}
    \hline
    QOP  & PLCK1 & PLCK2 &  Kept for the analysis \\
    \hline \hline
    1 & 341 & 411 & NO \\
    2 & 426 & 243 & YES \\
    3 & 124 & 50 & YES \\
    4 & 11 & 6 & YES \\
    6 & 20 & 24 & NO \\
    \hline
    \end{tabular}
    \label{tab:marz_results}
\end{table*}

\end{document}